# Dual Cross-Linked Hydrogels: Linear Rheology and Fractional Calculus Modeling

Agniva Dutta[a], Valeriy V. Ginzburg[b,*], Gleb Vasilyev[a], and Eyal Zussman[a,*]

[a] NanoEngineering Group, Faculty of Mechanical Engineering, Technion-Israel Institute of Technology, Haifa 3200003, Israel

[b] Department of Chemical Engineering and Materials Science, Michigan State University, East Lansing, Michigan 48224, USA

* Corresponding author emails: ginzbur7@msu.edu (VVG), meeyal@me.technion.ac.il (EZ)

## ABSTRACT

Hydrogels are increasingly recognized as a versatile platform for applications spanning from tissue engineering to soft robotics or flexible electronics. Recent efforts have focused on enhancing and tailoring their mechanical performance to meet application-specific demands. However, the intricate viscoelastic response of hydrogels remains challenging to capture using conventional phenomenological models. In this study, we prepared a series of tough dual crosslinked hydrogels - poly(methacrylamide-co-acrylic acid)-$Fe^{3+}$ and systematically tuned their mechanical properties by leveraging the salting-out effect. The viscoelastic behavior of the hydrogels was characterized under shear deformations, and a three-parameter fractional Maxwell Gel (FMG) model was constructed to quantitatively describe oscillatory shear, creep, and stress relaxation responses. The influence of salt concentration on each FMG parameter was analyzed and correlated with bulk mechanical performance. This framework provides a first step toward capturing the complex viscoelastic nature of the advanced hydrogels and lays the foundation for developing more comprehensive nonlinear constitutive models.

**For Table of Contents use only**

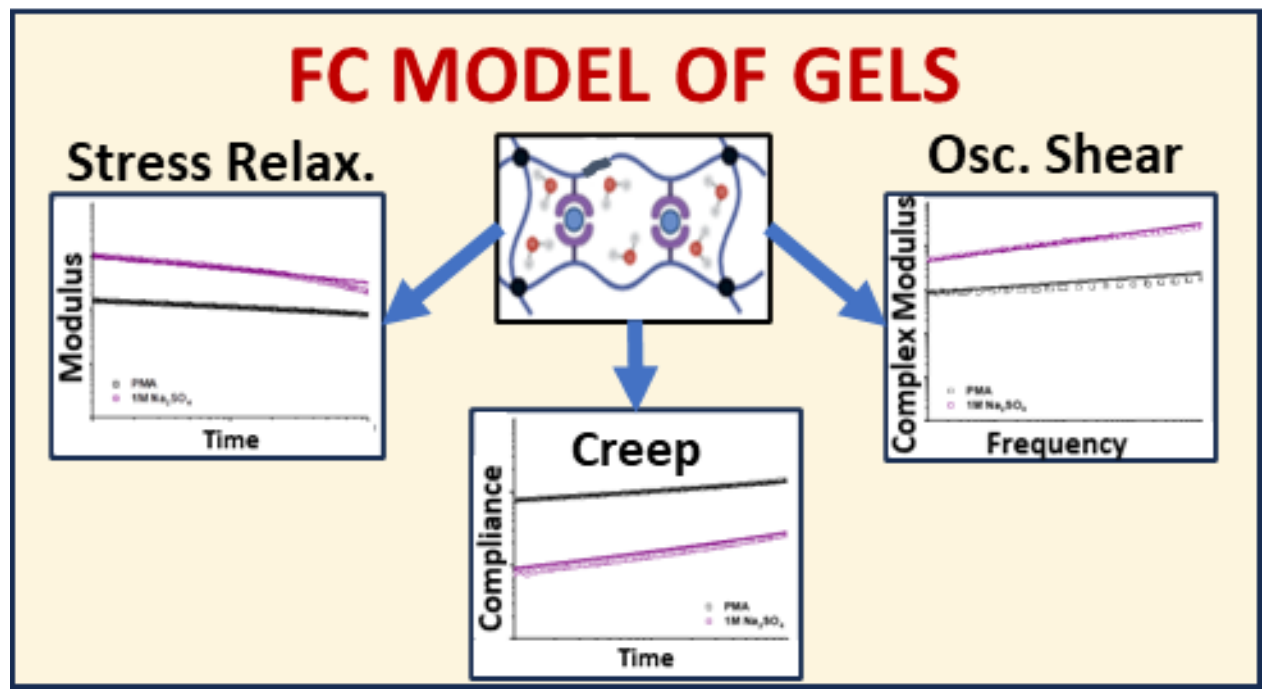

## 1. INTRODUCTION

Hydrogels, composed of crosslinked polymer networks swollen with water, closely resemble biological tissues and hold great promise for emerging technologies such as soft robotics,[1,2] wearable/flexible electronics,[3,4] sensors,[5,6] and energy storage devices.[7,8] A wide range of these applications demands hydrogels that can withstand substantial static and dynamic loads while resisting premature failure. Therefore, tailoring their mechanical properties to meet specific performance requirements is critical for practical implementation. Conventional hydrogels are inherently weak and brittle, primarily due to severe stress localization under load, often resulting in premature fracture. To overcome these limitations, various toughening strategies have been developed, primarily based on stress delocalization.[9] A widely adopted approach involves incorporating energy-dissipative motifs into the hydrogel network. These motifs preferentially break before the primary polymer backbone, dissipating applied energy and delaying catastrophic failure. Restoration of such motifs after breaking is essential to recover mechanical integrity, which is typically achieved using reversible physical bonds/interactions as crosslinkers.[10] Strategies beyond dissipation-based mechanisms can enhance material toughness by enabling the elastic absorption of energy. For example, slide-ring hydrogels incorporate figure-of-eight polyrotaxane crosslinks that can move freely along polymer chains, redistributing stress like pulleys.[11] This sliding mechanism effectively increases the chain length between crosslinks, improving extensibility. It is also possible for dense entanglements in swollen polyacrylamide hydrogels to slip due to reduced friction in water.[12] Unlike conventional covalent crosslinks, these mobile entanglements prevent premature fracture by distributing stress at crack tips.

Most reported hydrogels are viscoelastic, incorporating an additional design dimension, time, through their temporal structure. This temporal component is governed by the lifetime of cross-links, and even subtle changes in cross-linking kinetics can markedly influence bulk mechanical properties.[13] Notably, the mechanical response of soft materials often exhibits complexities that extend beyond the scope of classical phenomenological models, including the Maxwell, Kelvin-Voigt, or Zener (standard linear solid) frameworks. These traditional models, which combine Hookean springs and Newtonian dashpots, capture linear viscoelasticity but fail to account for the power-law scaling frequently observed in hydrogel material functions such as relaxation modulus and creep compliance. The Burgers model, which couples Maxwell with a Kelvin-Voigt element in a series, extends this frame to describe creep and stress relaxation. However, it remains constrained by discrete relaxation times and thus fails to capture broad relaxation spectra of hydrogels.[14,15] This deviation arises from the broad distribution of structural length scales and relaxation times inherent to the hydrogels. Consequently, a comprehensive understanding of both spatial and temporal design elements is essential for rationally tuning the hydrogel's mechanical properties.

In recent years, many problems dealing with systems having broad relaxation time distributions have been successfully addressed using the fractional calculus methods.[16–32] In particular, fractional viscoelastic models, which employ fractional derivatives instead of conventional first-order terms, provide a concise framework for describing complex mechanical responses. These models inherently capture power-law relaxation and broad time spectra, offering advantages over classical approaches such as generalized Maxwell model ("Prony series") which require multiple elements. Several of these studies were devoted to hydrogels. Jaishankar and McKinley used their Fractional Maxwell Gel (FMG) model to successfully describe the power-law stress relaxation in food hydrogels (caseinate and alginate).[22,25] Ruso and coworkers compared classical and fractional rheological models for hydroxyapatite-based composite hydrogels and found that the fractional models more accurately described their time-dependent viscoelastic behavior.[33] Miranda Valdez and coworkers similarly applied fractional viscoelastic models to conventional cellulose-based hydrogels, successfully capturing frequency-dependent viscoelastic responses across varying concentrations.[34] Lenoch et al. further combined fractional and generalized Maxwell models to describe the thermorheological behavior of poly(acrylic acid)-Fe(III) hydrogels, observing a transition in relaxation mechanism dominated by chain dynamics to the dissociation of coordination crosslinks as gelation progressed.[35] Several authors, most notably McKinley and co-workers,[26,36] and Ewoldt and co-workers,[37–39] highlighted the challenge of selecting the appropriate FC model for any particular material, and formulated a Bayesian approach to the FC model selection. Despite these advances, fractional viscoelastic models remain largely unexplored in the context of advanced, mechanically tunable hydrogels. This gap underscores the need for modeling approaches tailored to hydrogel systems with mechanical properties tunable across a broad spectrum.

In this study, we employed the Fractional Maxwell Gel (FMG) model to analyze the viscoelastic behavior of a dual cross-linked poly(methacrylamide-co-acrylic acid)-$Fe^{3+}$ (PMA) hydrogel.[40] The dual-crosslinking strategy integrates permanent covalent crosslinks with dynamic $Fe^{3+}$-acrylic acid coordination bond crosslinks, the latter breaking preferentially under external loading to dissipate energy.[41] To tune the mechanical properties across a broad spectrum, we exploited the salting-out effect by immersing the hydrogels in sodium sulfate solutions of varying concentrations. The mechanical response of both pristine and salt-treated PMA hydrogels was systematically examined under shear and tensile deformation, and a three-parameter FMG model was constructed to capture the oscillatory shear, creep, and stress relaxation behaviors, as well as the tensile modulus measured in uniaxial tensile tests.

## 2. MATERIALS AND METHODS

### *2.1 Materials.*

Methacrylamide (MAAm) and ammonium persulfate (APS) were purchased from Sigma Aldrich. Acrylic acid (AAc) and N,N'-Methylenebisacrylamide (MBAA) were purchased from Merck, and N,N,N',N'-tetramethylethylenediamine (TEMED) from Acros Organics. Ferric nitrate nonahydrate [$Fe(NO_3)_3 \cdot 9H_2O$] was purchased from Fisher Scientific, and anhydrous Sodium sulfate ($Na_2SO_4$) from Bio-lab Ltd. All chemicals were used as received without further purification.

### *2.2 Preparation of the Poly(MAAm-co-AAc)-$Fe^{3+}$ hydrogel*

The dual crosslinked hydrogel was prepared through a conventional three-step procedure.[40,41]

Step 1 - Synthesis of chemically crosslinked hydrogel:

MAAm (2.4 M), AAc (10 mol% relative to MAAm), and MBAA (0.06 mol% relative to the total monomer content) were dissolved in deionized water. The solution was purged with nitrogen gas for 15 min to remove dissolved oxygen. Subsequently, initiator APS (1 wt.% relative to total monomer weight) and accelerator TEMED (20 μL) were added under constant stirring. The resulting solution was slowly poured into a Teflon mold and allowed to polymerize at room temperature for 12 hours to obtain chemically crosslinked hydrogels.

Step 2 - Ionic crosslinking with $Fe^{3+}$:

The chemically crosslinked hydrogels were immersed in a 0.06 M $Fe(NO_3)_3$ solution for 24 h to introduce ionic crosslinking via $Fe^{3+}$ coordination with carboxylic groups.

Step 3 - Post-treatment:

The hydrogel samples were immersed in a large volume of deionized water for 48 h, with the water being replaced periodically to remove excess $Fe^{3+}$. Finally, the hydrogels were treated with sodium sulfate solution of varying molarities to get the final dual crosslinked hydrogels.

### *2.3 Characterization*

The mechanical properties of the hydrogels under shear deformation were characterized using a Discovery DHR-2 rheometer (TA Instruments, USA) equipped with a parallel plate geometry (20 mm diameter, 1.4-1.7 mm gap). To prevent slippage during testing, 240-grit waterproof sandpaper was affixed to the plate surfaces. Frequency sweep experiments were carried out at room temperature (T = 25 °C) within the linear viscoelastic region, determined by preliminary amplitude sweep test. (See Figure S1 in the Supporting Information for more details). The strain amplitude was 0.04%, and the frequency range was between 0.1 and 100 rad/s for the oscillatory shear. Stress-relaxation measurements were performed at 25 °C, by applying an instantaneous strain and recording the stress decay over a 12 min period. The values of applied strain belonged to the linear viscoelastic region and varied in the range 0.06 - 0.6% depending on the $Na_2SO_4$ molarity. Creep tests were carried out by continuous shearing of samples at a constant stress for 5 min at 25 °C. The values of applied stress varied in the range 200 - 400 Pa depending on the $Na_2SO_4$ molarity.

The tensile tests and their results were summarized in an earlier publication,[40] so here we provide only a brief summary. The tensile measurements were performed at room temperature using a Zwick/Roell Z050 Universal Testing Machine. Samples were coated with thin layer of silicon oil prior to the measurement to prevent water loss. For tensile testing, hydrogel sheets (~ 1 mm thick) were cut into rectangular strips (~ 40 mm in length and ~ 4 mm in width) and subjected to a grip-to-grip separation of ~10 mm. The extension rate was set at 100 mm/min.

### *2.4 Modeling – the Fractional Maxwell Gel (FMG) for Linear Viscoelasticity*

To model the linear viscoelasticity of the gels, we utilize the Fractional Maxwell Gel (FMG), schematically shown in Figure 1.

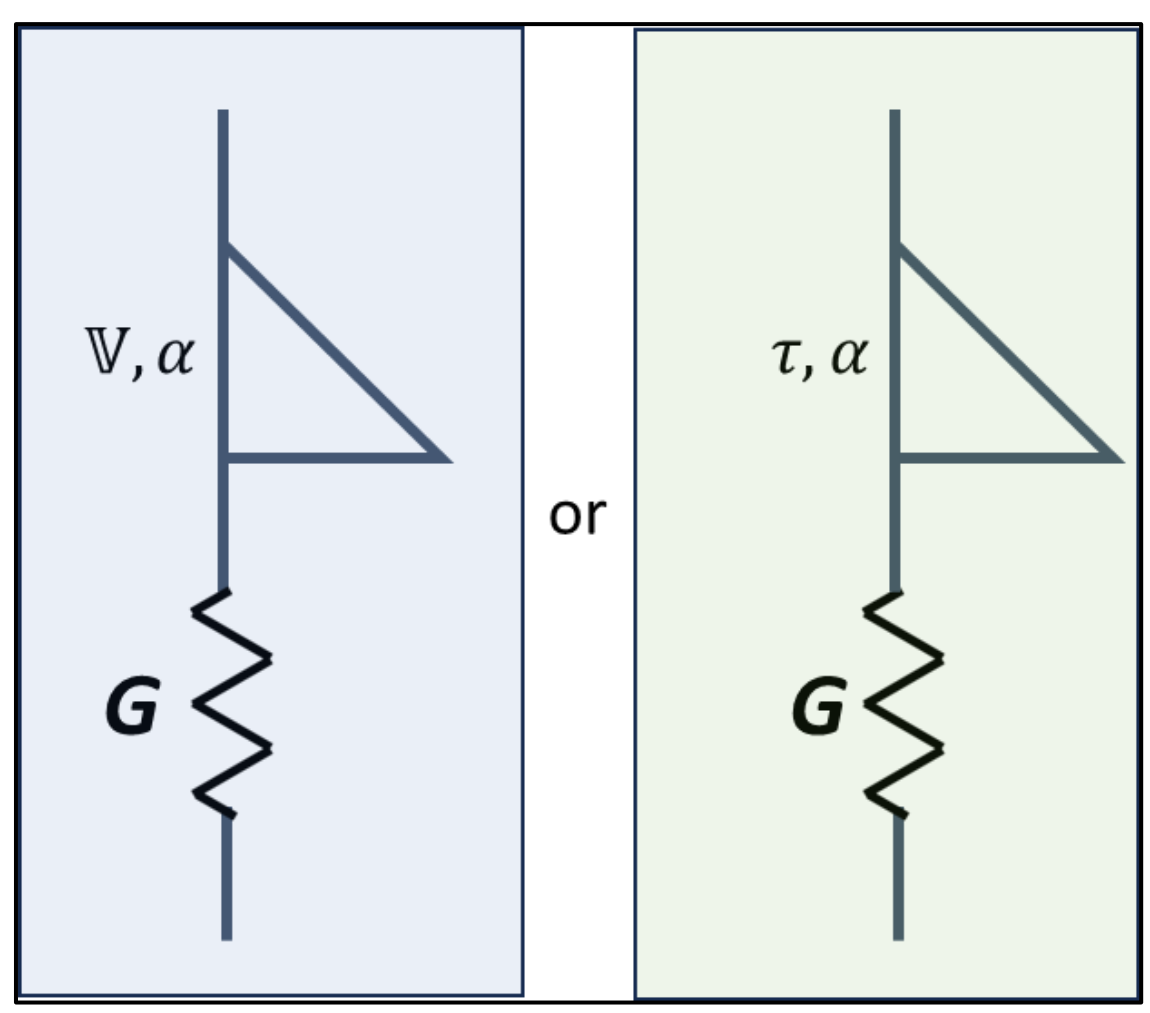

Figure 1. Schematic representation of the Fractional Maxwell Gel (FMG). The triangle represents a Scott Blair element connected in series with a Hookean dashpot (zigzagged line). The SB element can be parameterized using a quasi-property $\mathbb{V}$ (left panel) or a characteristic time $\tau$ (right panel). See text for more details.

As proposed by McKinley and co-workers,[21,23,25,26,28] the linear viscoelasticity of gels and other complex fluids can be described based on the power-law Scott Blair[42–44] (SB) elements. Each individual SB element (represented as a triangle in Figure 1) can be thought of as a representation of a material with an infinitely broad relaxation spectrum. In particular, the linear viscoelastic storage and loss moduli of an SB element are given by,

$$G'(\omega) = \mathbb{V}\, cos\left[\frac{\pi\alpha}{2}\right]\omega^{\alpha} \tag{1a}$$

$$G''(\omega) = \mathbb{V}\, sin\left[\frac{\pi\alpha}{2}\right]\omega^{\alpha} \tag{1b}$$

Here, $\mathbb{V}$ and $\alpha$ are the so-called "quasi-property"[21,26] and the power-law exponent of the SB element, respectively. Note that for $\alpha = 0.5$, $G' = G''$ for all frequencies $\omega$. Thus, SB element with $\alpha = 0.5$ describes the "ideal Winter-Chambon" gel.[45–48] The SB elements with $0 < \alpha < 0.5$ describe the "solid-like" gels, with $\alpha = 0$ corresponding to the classical Hookean spring. The SB elements with $0.5 < \alpha < 1$ describe the "liquid-like" gels, with $\alpha = 1$ corresponding to the classical Maxwell dashpot. The complex modulus, stress relaxation, and creep compliance of a single SB are given by,[26]

$$|G^*(\omega)| = \mathbb{V}\omega^{\alpha} \quad (2a)$$

$$G(t) = \mathbb{V}\frac{t^{-\alpha}}{\Gamma(1-\alpha)} \quad (2b)$$

$$J(t) = \mathbb{V}^{-1}\frac{t^{\alpha}}{\Gamma(1+\alpha)} \quad (2c)$$

In the FMG, an SB element is connected in series with a Hookean spring characterized by the shear modulus, $G$. This shear modulus corresponds to the plateau storage modulus at high frequency or the relaxation modulus at low times. The storage and loss moduli of the FMG are given by,

$$\frac{G'(\omega)}{G} = \frac{(\omega\tau)^{\alpha}\cos\left(\frac{\pi\alpha}{2}\right)+(\omega\tau)^{2\alpha}}{1+2(\omega\tau)^{\alpha}\cos\left(\frac{\pi\alpha}{2}\right)+(\omega\tau)^{2\alpha}} \quad (3a)$$

$$\frac{G''(\omega)}{G} = \frac{(\omega\tau)^{\alpha}\sin\left(\frac{\pi\alpha}{2}\right)}{1+2(\omega\tau)^{\alpha}\cos\left(\frac{\pi\alpha}{2}\right)+(\omega\tau)^{2\alpha}} \quad (3b)$$

The creep compliance, $J(t)$, can likewise be easily described in elementary and special functions,[21]

$$J(t) = \frac{1}{G}\left[\frac{1}{\Gamma(1+\alpha)}\left(\frac{t}{\tau}\right)^{\alpha}+1\right] \quad (4)$$

where $\Gamma(x)$ is the gamma-function.

The stress relaxation, *G(t)*, can be obtained as an inverse Fourier transform of the complex modulus $G^*(\omega) = G'(\omega) + iG''(\omega)$. Although it cannot be expressed in elementary functions, *G(t)* can be described using the so-called Mittag-Leffler special function, $E_a(x)$,[25]

$$G(t) = GE_\alpha\left[-\left(\frac{t}{\tau}\right)^\alpha\right] \quad (5a)$$

where,

$$E_a(x) = \sum_{k=0}^{\infty} \frac{x^k}{\Gamma(ak+1)} \quad (5b)$$

Thus, if a complex system (such as a polymer gel) can be described by a single FMG, the three independent experiments (stress relaxation, creep, and frequency sweep) can be captured using only three parameters: $G$, $\tau$, and $\alpha$. However, one might find that the linear viscoelasticity tests all fall into the time/frequency range where the experimental results are fully captured by a single power-law. In this case, one cannot independently estimate $G$ and $\tau$, but only the quasi-property $\mathbb{V} = G\tau^\alpha$. To estimate the plateau modulus, $G$, and the elementary time, $\tau$, separately, one needs to provide some additional information. Here, we propose to use the tensile data for the estimate of $G$ and then regress $\tau$ accordingly, as discussed below.

### *2.5. Modeling -- The Model Parameter Optimization Procedure*

The parameter optimization is performed in two steps.

Step 1 – Scott Blair Analysis. From the linear viscoelastic shear data, we compute the power-law exponent, $\alpha$, and the quasi-property, $\mathbb{V}$. For this two-parameter fitting, we attempt to minimize the following objective function characterizing the difference between the experimental data and the model fit:

$$\Omega = w_{SR}\Omega_{SR} + w_C\Omega_C + w_{FS}\Omega_{FS} \quad (6a)$$

Here, the indices SR, C, and FS stand for "stress relaxation", "creep", and "frequency sweep", respectively; $w_i$ are the weights associated with each of the three experiments. We choose $w_{CR} = w_C = w_{FS} = 0.333$. The individual objective functions $\Omega_{SR}$, $\Omega_C$, and $\Omega_{FS}$ are given by,

$$\Omega_{SR} = \frac{1}{N_{SR}} \sum_{k=1}^{N_{SR}\Sigma} \left| log \frac{G[k]_{exp}}{G_{model}(t[k])} \right| \tag{6b}$$

$$\Omega_C = \frac{1}{N_C} \sum_{k=1}^{N_C\Sigma} \left| log \frac{J[k]_{exp}}{J_{model}(t[k])} \right| \tag{6c}$$

$$\Omega_{FS} = \frac{1}{N_{FS}} \sum_{k=1}^{N_{FS}\Sigma} \left| log \frac{|G^*_{exp}|[k]}{|G^*_{model}|(\omega[k])} \right| \tag{6d}$$

Here, $k = 1, N_i$, $i = SR, C,$ or $FS$, corresponds to the index of a specific data point in the experimental array, and $N_i$ are the array lengths (numbers of points); $Y_{exp}[k]$ are the experimental data, with Y = $G$ (for $SR$), $J$ (for $C$), and $|G^*|$ (for $FS$). The *log* operator means a base-10 logarithm. The complex modulus magnitude is given by,

$$|G^*| = \sqrt{(G')^2 + (G")^2} \tag{7}$$

Note that we use the complex modulus, $|G^*|$, rather than the storage and loss moduli $G'$ and $G"$ because for the power-law spectrum characteristic of the gels, the three functions are proportional and we are free to choose any one of them for our analysis. The model values $G_{model}(t)$ are given by equations 2b, $J_{model}(t)$ by equation 2c, and $|G^*_{model}|(\omega)$ by equation 2a.

The ranges used for optimization are 1 – 100 s for the creep and stress relaxation experiments, and 0.1 – 100 rad/s for the oscillatory shear experiments. The parameter optimization is then done in Microsoft Excel 2019 using the Generalized Reduced Gradient (GRG) procedure.

Step 2. Once $\mathbb{V}$ and $\alpha$ are regressed for each gel, we then attempt to estimate $G$ and $\tau$ as follows. For each salt concentration, $c$, we use the tensile measurement of the modulus, $E(c)$, and compute the corresponding shear modulus, $G_t(c) = E(c)/3$, where the index "t" stands to reflect the fact that the shear modulus was estimated based on the tensile stress-strain test.

We assume that $G_t(c)$ is a reasonable approximation for the plateau modulus, $G$, in which case the relaxation time is given by,

$$log(\tau) = \frac{log(\mathbb{V}) - log(G)}{\alpha} \approx \frac{log(\mathbb{V}) - log(G_t(c))}{\alpha} \quad (8)$$

Here, $\tau$ is in s, $G$ (or $G_t(c)$) is in Pa, and $\mathbb{V}$ is in Pa·s$^{\alpha}$; *log* means base-10 logarithm. To smooth out the variations due to possible error in $\alpha$, we stipulate that the relaxation time is a quadratic function of $c$ (where $Na_2SO_4$ concentration $c$ is varied between 0 and 1M),

$$log\big(\tau(c)\big) = A + Bc + Cc^2 \quad (9)$$

We then minimize the following objective function (with respect to A, B, and C),

$$W = \frac{1}{5}\sum_{i=1}^{5}\left|\left\{log\mathbb{V} - \alpha log\big(\tau(c_i)\big)\right\} - log\big(G_t(c_i)\big)\right| \quad (10)$$

The above procedure ensures that the relaxation time dependence on $c$ is a smooth function, while the plateau modulus is close to (within an order of magnitude) the value estimated based on the stress-strain curves.

The two-step procedure could, in principle, be replaced by a single-step parameter optimization where the full FMG equations (equations 3a, 3b, and 7 for the complex modulus; equation 4 for the creep compliance, and equations 5a – 5b for the relaxation modulus) are used instead of the SB equations 2a – 2c. In splitting the process into two parts, we reduce the chances of falling into a “wrong” local minimum of the objective function in the phase space. The two approaches should result in similar answers as long as the Deborah number, $t/\tau$ (for stress relaxation and creep), or $1/\omega\tau$ (for oscillatory shear), is sufficiently large. We will see that this condition is fully satisfied for all our gels.

## 3. RESULTS AND DISCUSSION

### *3.1 The Two Networks.*

In Figure 2, we schematically show the structure of the prepared hydrogels. There are two types of effective crosslinks in the hydrogels - the covalent bonds (black dots) and the ionic physical crosslinks (blue circles with purple semicircles). The overall viscoelastic response of the hydrogel depends on the relative density of the two types of crosslinks. As discussed in our previous papers,[40,49] the addition of salt causes the phase separation of excess water, making the hydrogels stiffer. Increasing the salt concentration leads to increased phase separation ("syneresis"), and a substantial increase in the gel modulus and strength.

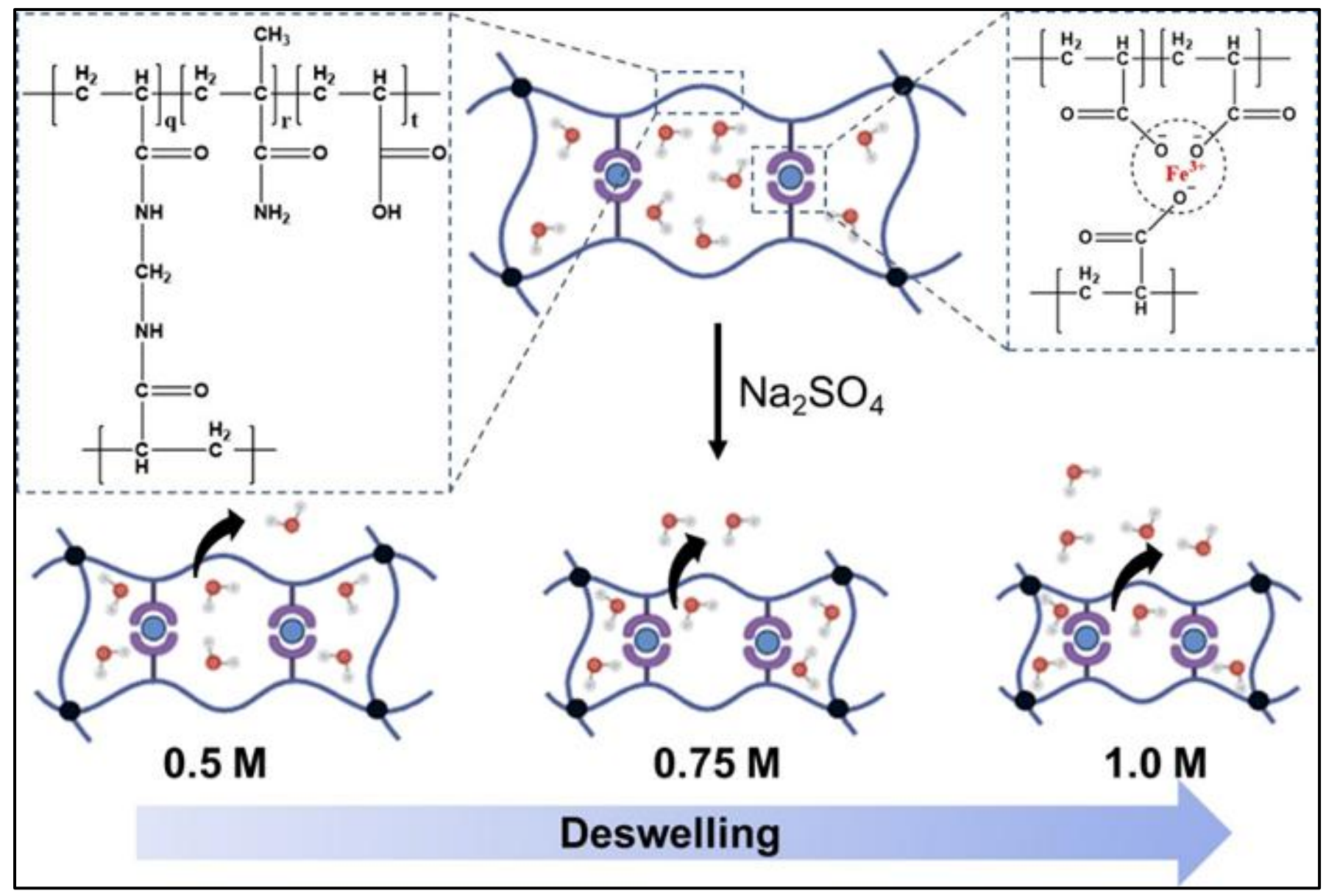

Figure 2. The chemistry of the hydrogels, two types of crosslinks, and the syneresis occurring due to the addition of the sodium sulphate salt.

As discussed above, we attempt to model our systems using the FMG approach. The linear viscoelasticity experiments result is nearly power-law dependences for *G(t)*, *J(t)*, and $|G^*(\omega)|$, indicating that they can be well-approximated by the Scott Blair element. The Hookean element is then parameterized based on the modulus estimated from the tensile experiments. We performed the FMG parameterization for the five PMA gels with varying salt content. The results are discussed below.

### *3.2 Stress Relaxation*

The stress relaxation (SR) experiment describes the time-dependent shear modulus, *G(t)*, as a function of time after one applies an instantaneous strain and records the stress decay; the time-dependent shear modulus is the ratio of the time-dependent stress to the (constant) strain. The experimental data and corresponding SB model fits are shown in Figure 3a for the no-salt PMA and 1M $Na_2SO_4$ PMA gels, and in Figure 3b for all the intermediate salt concentrations. (The SB model parameters are summarized in Table 1 at the end of this section).

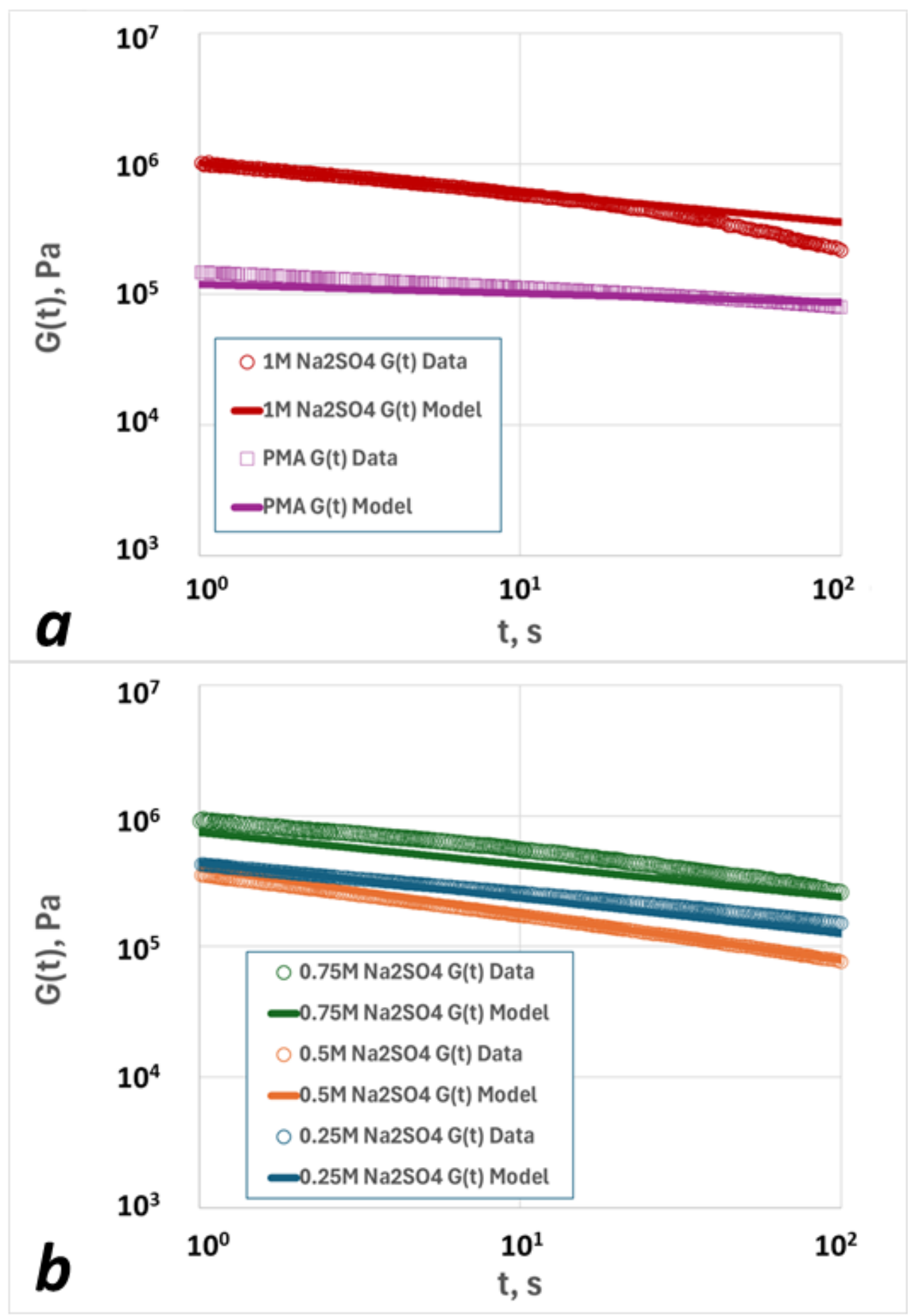

Figure 3. Stress relaxation experiments data (symbols) and SB model fits (solid lines). (a) Pure PMA with no salt (purple), and PMA with 1M $Na_2SO_4$ (red) (b) PMA with 0.25M (blue), 0.5M (orange), and 0.75M (green) of $Na_2SO_4$.

One can draw several conclusions from the figures. First, the SB fits provide very good descriptions of the relaxation data, at least within the time range considered in our analysis (1 – 100 s). The high-salt 1M $Na_2SO_4$ system is one exception – at large times, it shows a deviation from the model prediction, relaxing faster than expected. This can perhaps be explained by some weakly nonlinear effects and/or by spontaneous break-up of various clusters – the processes not captured within the SB (or FMG) framework. Second, the stress-relaxation curves show a fairly complex behavior, especially for the intermediate (0.25M, 0.5M, and 0.75M) salt concentrations. The 0.5M gel exhibits lower modulus than the 0.25M one, breaking with the trend of monotonic modulus increase with the salt concentration. However, the power exponent $\alpha$ has a maximum for the 0.5M gel, so one can safely assume that the plateau modulus is indeed a monotonic function of $c$, and this could potentially be seen if one can resolve the limit of extremely small times (microseconds or less).

In the limit of very small times, the modulus $G(t)$ reaches its plateau value, which is expected to be close to the shear modulus evaluated from the tensile experiments. It is illustrated in the Supporting Information.

### *3.3 Creep*

The creep test measures the time-dependent deformation of the gel sample subjected to a constant load; the deformation is then converted to creep compliance, $J$. The experimental and SB fitting results for the creep test are shown in Figures 4a and 4b (similar groupings as in Figure 3). In this case as well, the SB model fits well with the data. The SB parameters are the same as for the stress relaxation experiments (Table 1).

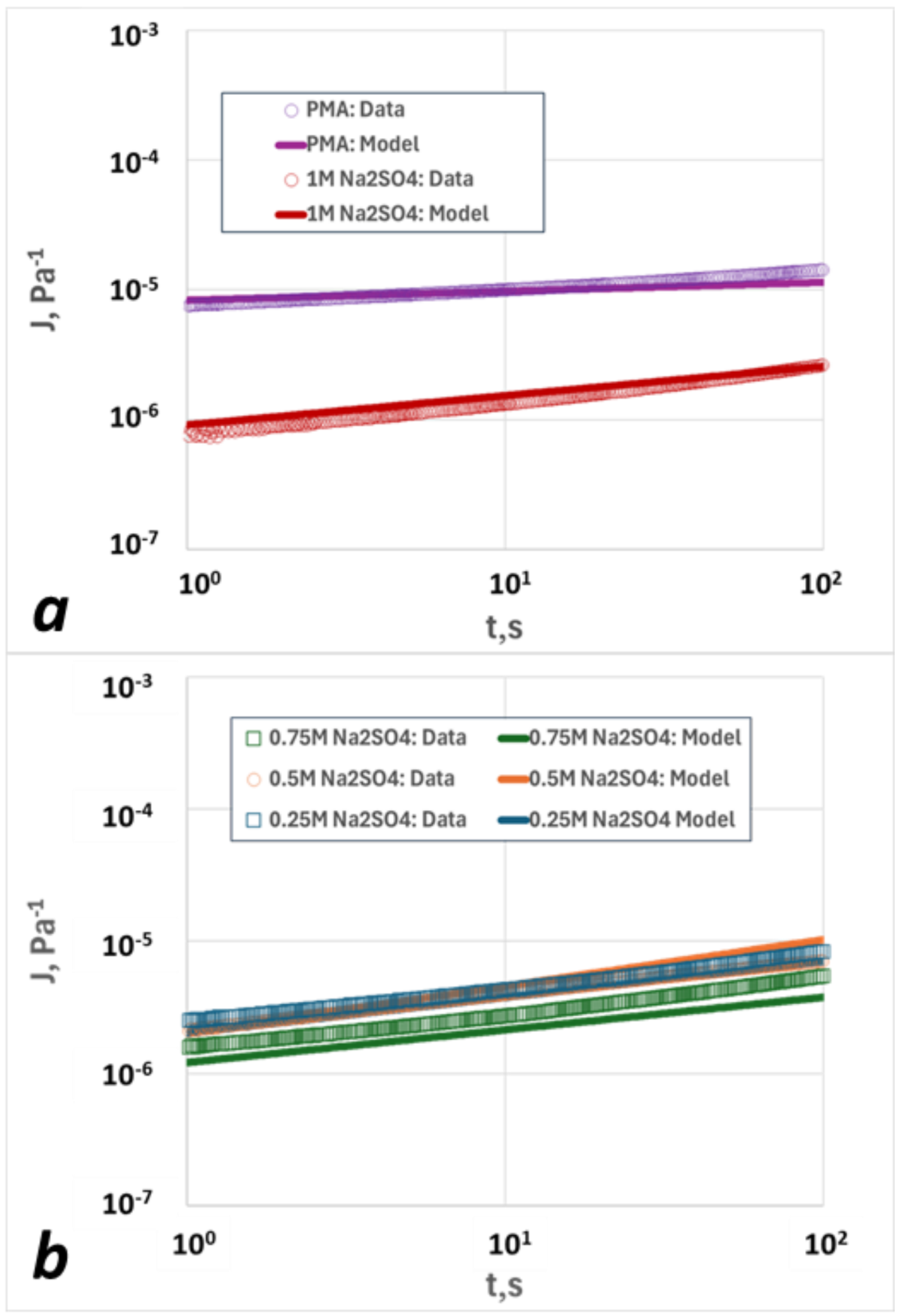

Figure 4. Creep experiments data (symbols) and SB model fits (solid lines). (a) Pure PMA with no salt (purple), and PMA with 1M $Na_2SO_4$ (red) (b) PMA with 0.25M (blue), 0.5M (orange), and 0.75M (green) of $Na_2SO_4$.

Creep compliance is in some ways the inverse of the modulus, so the higher-compliance materials are the softer ones, and the lower-compliance ones are the stiffer gels. Again, as expected, the intermediate-salt-concentration (0.25M and 0.5M) gels exhibit very similar behaviors, as do the higher-salt-concentration ones (0.75M and 1M).

### *3.4 Oscillatory Shear Modulus*

To fit our model, we did not use the storage and loss moduli separately, but rather the overall complex modulus amplitude, |G*| (equation 6). The results are plotted in Figures 5a and 5b.

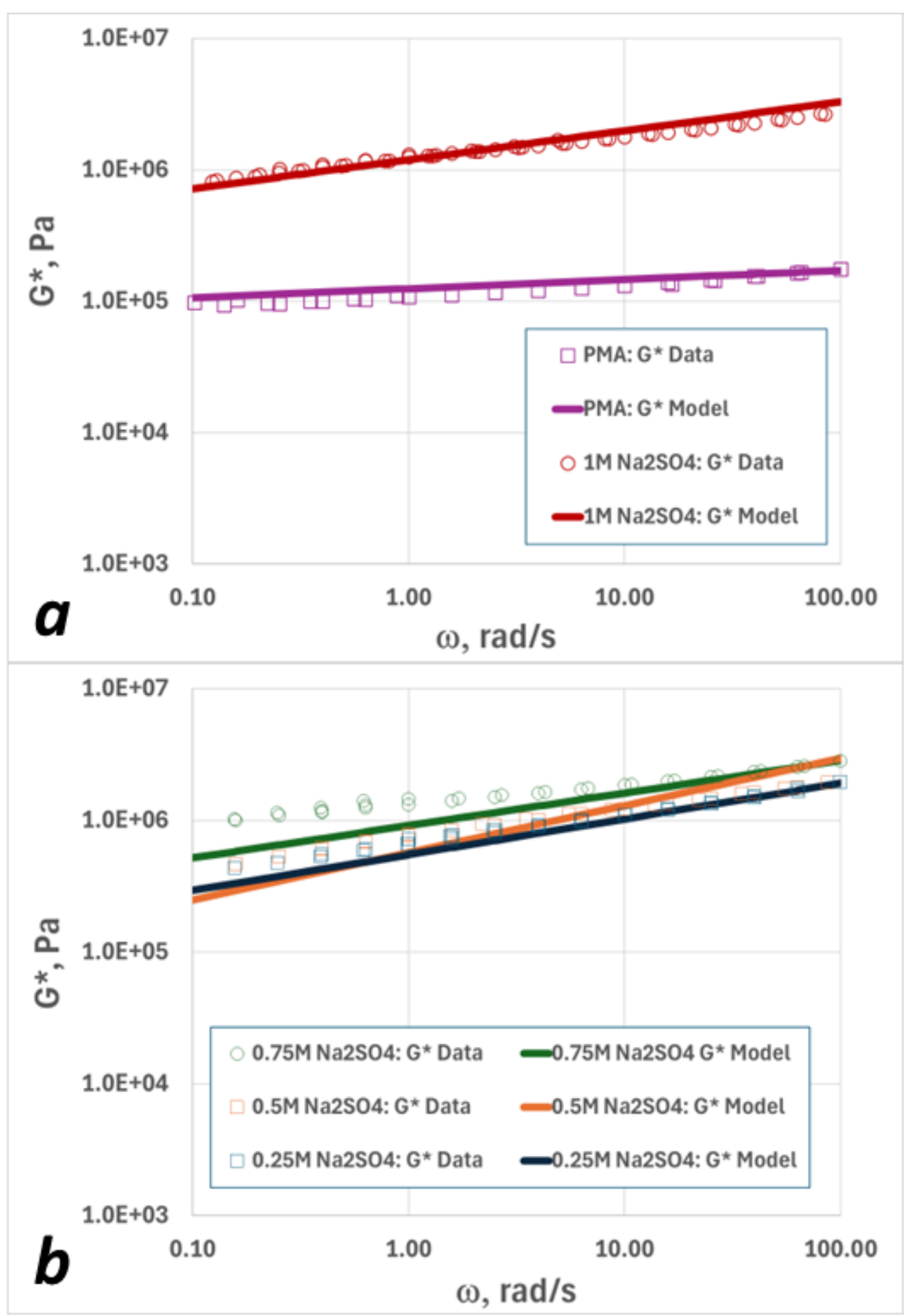

Figure 5. Frequency sweep experiments data (symbols) and SB model fits (solid lines). (a) Pure PMA with no salt (purple), and PMA with 1M $Na_2SO_4$ (red) (b) PMA with 0.25M (blue), 0.5M (orange), and 0.75M (green) of $Na_2SO_4$.

The comparison between the model and the SB fits shows reasonable agreement, except for the 0.75M salt gel at low frequencies. The dependence of the complex modulus amplitude on the frequency is approximately – but not exactly – a power-law (which is a signature of a Scott Blair element). The dependence of $|G^*|$ on the salt concentration is monotonic, with the gel rigidity increasing upon the addition of the salt, as one would expect.[40,49] The model parameters are the same as for the other two cases (Table 1).

### *3.5 Salt Concentration Dependence of the Model Parameters*

The FMG model parameters are summarized in Table 1 for all five gels considered here, and their dependence on the salt content is depicted in Figures 6 and 7a-b.

Table 1. FMG model parameters for different salt concentrations: logarithm of the shear modulus, G; power-law exponent, a; logarithm of the relaxation time, t; the objective function, W. See text for more details.

| | **PMA** | **0.25M $Na_2SO_4$** | **0.5M $Na_2SO_4$** | **0.75M $Na_2SO_4$** | **1M $Na_2SO_4$** |
|---|---|---|---|---|---|
| $\alpha$ | **0.07** | **0.27** | **0.36** | **0.25** | **0.22** |
| **log(V)** | **5.1** | **5.7** | **5.76** | **6.0** | **6.1** |
| **log(G, Pa)** | **5.4** | **6.4** | **6.68** | **7.1** | **8.0** |
| **log($\tau$, s)** | **-4.2** | **-2.4** | **-2.6** | **-4.7** | **-8.9** |
| $\Omega$ | **0.062** | **0.048** | **0.061** | **0.078** | **0.055** |

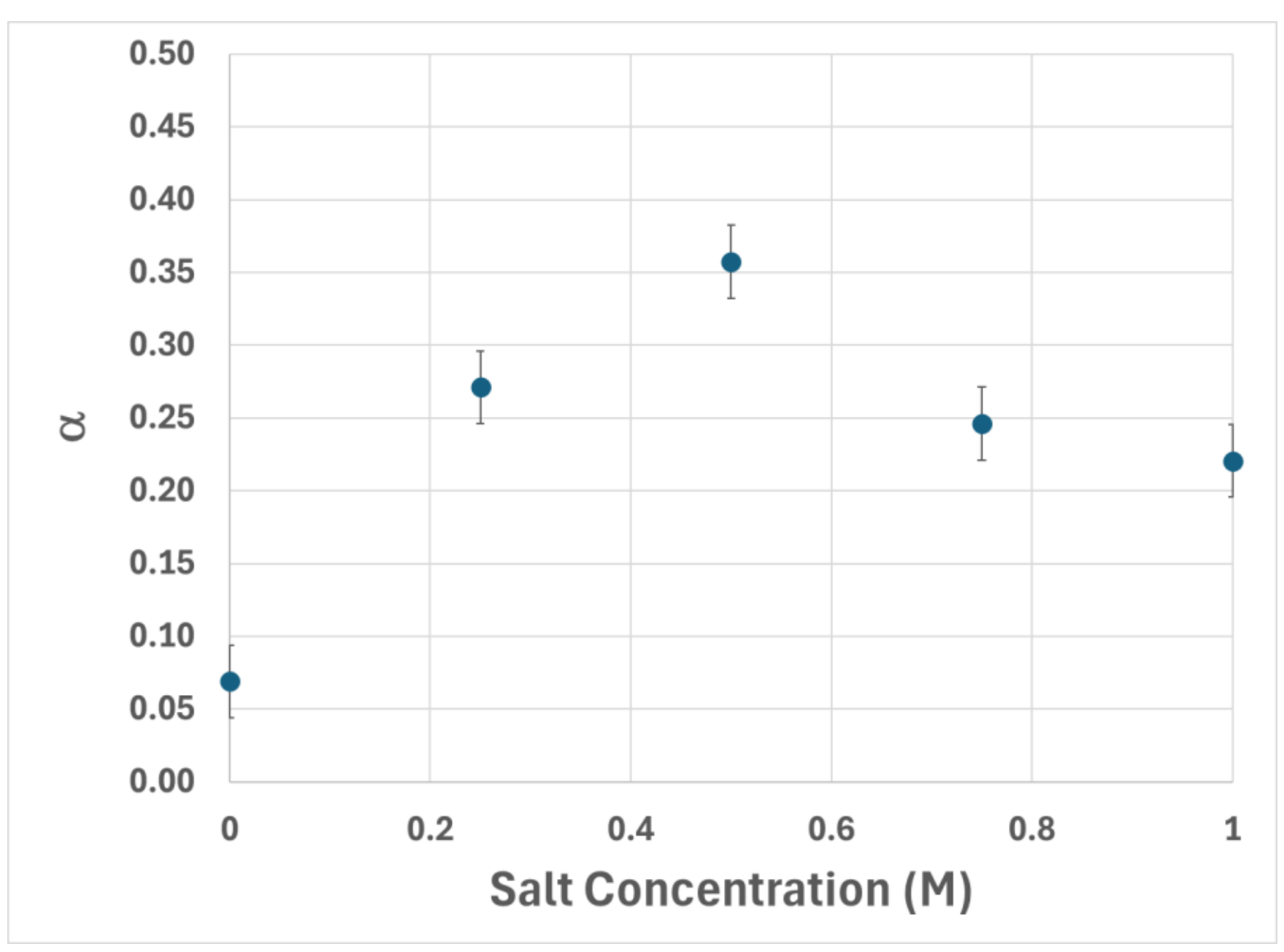

Figure 6. Salt concentration dependence of the power-law exponent, α. The Y-error bars are 0.025.

The power-law exponent $\alpha$ shows a non-monotonic dependence on the salt concentration. The exponent is very small (~0.07) for the no-salt system, increases to ~0.35 for the intermediate salt concentration, and then decreases for the high-salt systems. This indicates that the no-salt system has the nearly rate-independent elasticity, similar to an ideal elastic material; the intermediate-salt systems are relatively close to the "ideal gel" ($\alpha = 0.5$), which is "equally" liquid and solid; further increase in the salt concentration makes the material "more solid" again. As expected, this behavior resembles the transitions from "liquid-like" to "solid-like" gels upon increasing crosslinking.[45,46,50,51]

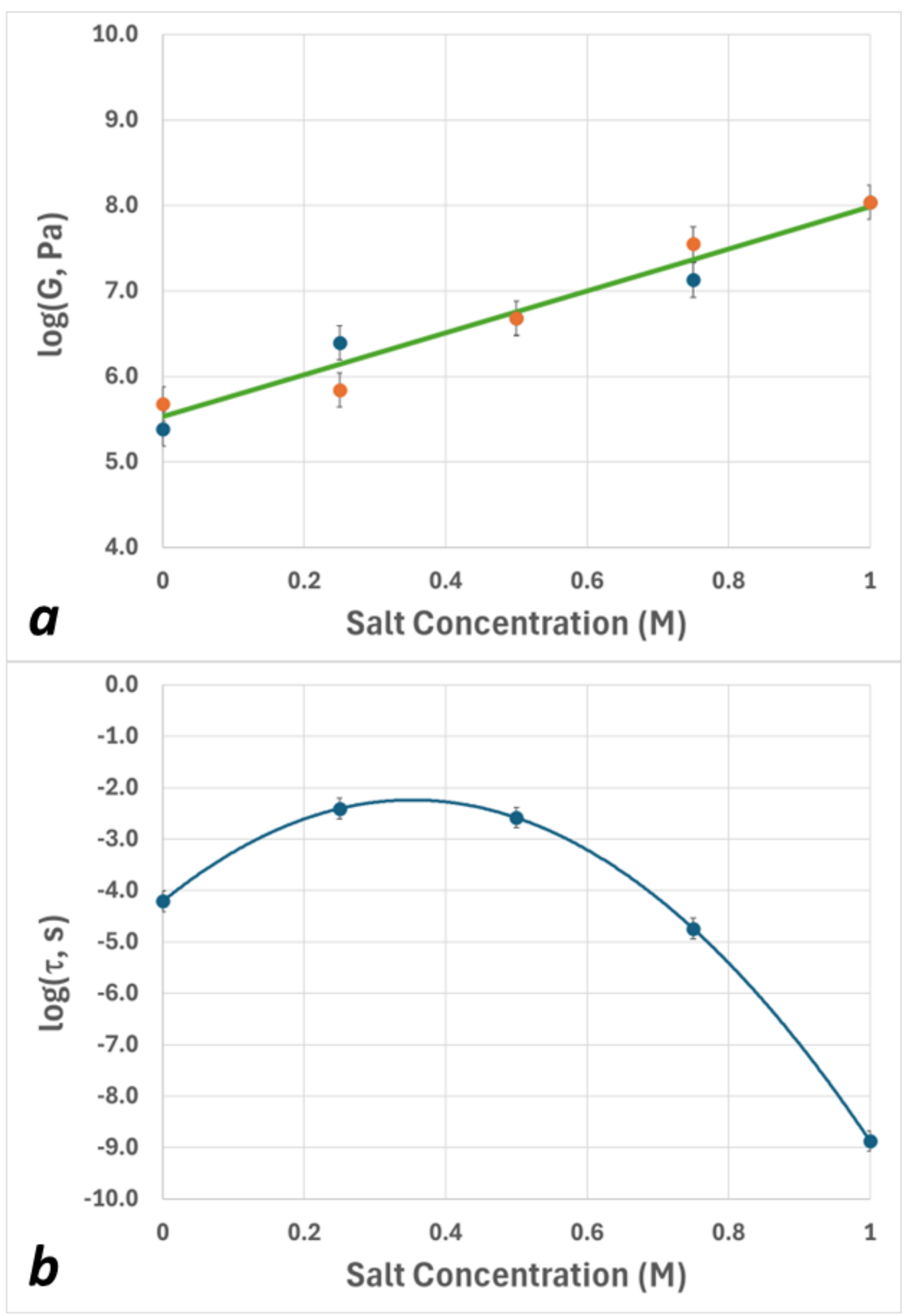

Figure 7. (a) Logarithm of the estimated plateau modulus, G (blue circles), the modulus from tensile measurements, $G_t(c)$ (orange circles), and a linear fit to both sets combined (green line). The Y-error bars are 0.2. The dashed line is guide to eye. (b) Logarithm of the estimated characteristic time, $\tau$ (blue circles), and the quadratic fit. The Y-error bars are 0.2.

The modulus magnitude, *G* (Figure 7a) increases with rising salt concentration, consistent with previous observations where stiffness and strength were enhanced as a function of salt concentration.[40] The increase in modulus, which reflects the effective crosslinking density, can be attributed to salt-induced deswelling that forces polymer chains towards high energy conformations. Since the mechanical properties of hydrogels are often governed by the lifetime of physical cross-links, systems with slow binding/unbinding kinetics are expected to exhibit higher strength, whereas fast exchange rates yield softer gels. Interestingly, in PMA hydrogels, while the modulus increased with salt concentration, the "elementary" relaxation time $\tau$ exhibits a non-trivial, non-monotonic dependence on the salt concentration (Figure 7b), given by equation 9, with A = -4.21, B = 11.16, and C = -15.83. Note that for all gel systems considered here, $\tau < 10^{-2}$ s, and thus for the range of times considered in our experiments ($t \sim 1 - 100\ s$ or $\omega^{-1} \sim 0.01 - 10\ s/rad$), the Deborah number is always greater than unity, and the difference between the full FMG model and its SB asymptotic is small enough to be neglected, as we originally supposed.

The observed trends can be tentatively explained as follows. At low salt concentrations, the hydrogels are fairly swollen. The crosslink density is low, the ionic crosslink lifetime is relatively small, and the dispersity in the crosslink lifetimes is narrow. As the salt concentration is increased, the water is expelled, leading to significant increase in the crosslink/ionic bond density. The crosslink lifetimes also begin to increase, and so does the lifetime distribution. At some point, however, the crosslinks are likely to become sterically constrained, leading to the reduction in the lifetime and the lifetime dispersity (while the crosslink density still continues to increase). With the crosslink density proportional to the plateau modulus $G$, the crosslink lifetime related to the characteristic time $\tau$, and the crosslink lifetime dispersity related to the power exponent $\alpha$, the proposed picture is consistent with the FMG analysis.

### 3.6 Discussion

The fractional calculus (FC) analysis of several PMA-based ionic gels prepared with different salt concentrations shows an interesting and complex rheological behavior. The characterization included three separate measurements – stress relaxation (SR), creep (C), and oscillatory shear (OS). Within some relatively well-defined time (for C and SR) or frequency (OS) ranges, FC provides a reasonably good description of all three tests. There have been very few examples where a single fractional model can capture these tests simultaneously (see, e.g., Aime et al.[52]). Thus, the current analysis offers further support for the idea that FC models can provide a parsimonious and physically meaningful description of the linear rheology of complex fluids (including hydrogels).

In our analysis, we found that all the data from the stress relaxation, creep, and oscillatory shear experiments are consistent with a power-law dependence and thus can be fully described by the Scott Blair model with two parameters – the power-law exponent $\alpha$ and the "quasi-property" $\mathbb{V}$. However, the SB model does not have a finite "plateau modulus", expecting, e.g., $|G^*|$ to grow to infinity as $\omega \rightarrow \infty$. We thus hypothesized that the modulus obtained from stress-strain tests can serve as a good proxy for the plateau modulus – or at least a reasonable lower bound for it.

The FMG parameterization represents a simple and parsimonious way of capturing the broad distribution of relaxation times inherent to gels. However, implementing FMG is other models (e.g., Finite Element Analysis [FEA] solvers) is often not straightforward. In that case, one can revert to traditional spectral methods such as Prony series. Song et al.[26] (and many other authors) provide an analytical expression for the continuous relaxation spectrum $H(\tau)$ which can be used to parameterize FEA or analytical visco-elasto-plastic modeling.

There are a number of other questions that are not fully answered by the current model. For example, the stress relaxation accelerates at long times, and this behavior is not captured within a single FMG. Adding new elements would probably improve the agreement between model and experiment, but at the cost of introducing new parameters with no clear physical meaning. Is the enhanced relaxation at longer times due to the spontaneous bond-breaking not tied to the built-in strain? Or are there nonlinear effects in play? Likewise, are the discrepancies between the model and experiment in the low-frequency region for the OS experiment due to the failure of the model, potential nonlinearity, or some other factors? Will there be any additional modes observed at high frequencies and/or short times? We will be exploring these topics in the future.

Finally, it is important to note that the linear tests – as important as they are – are only the first step in the evaluation of the hydrogels and prediction of their properties and application performance. Previously, cyclic tensile loading–unloading tests demonstrated that the hysteresis area, representing the energy dissipation relative to toughness, increased markedly with salt concentration.[40] Together, we can tentatively stipulate that the ionic crosslinks are forced into more "rigid", higher-energy conformations in the higher-salt-concentration gels. Accordingly, at low deformations, the barrier to escape those conformations becomes smaller, and the stress is transferred to weaker physical crosslinks, making them more prone to rupture, reducing their lifetime. Toughness could be defined as the sum of dissipated and elastically stored energy. As observed in cyclic tensile tests, higher salt concentrations increased the relative contribution of energy dissipation compared to elastic storage. The FC models parameterized here should be considered a stepping stone *en route* to building a comprehensive nonlinear dynamic stress-strain model.[22,24,25,49]

## 4. CONCLUSIONS

We characterized the linear rheology of PMA-based ionic hydrogels and modeled it using fractional calculus approach. A simple two-parameter Scott Blair model is shown to successfully capture the oscillatory shear, creep, and stress relaxation experiments. By including the data from tensile stress-strain tests, we generalize our analysis to the three-parameter Fractional Maxwell Gel (FMG) model, thus imposing the higher bound on the plateau modulus. We investigated the dependence of the model parameters on the salt concentration and found interesting and non-trivial dependencies that will be investigated in the future. The calculated FMG model parameters will be used to develop the dynamic nonlinear stress-strain models.

## 5. ASSOCIATED CONTENT

Supporting Information

The Supporting Information is available free of charge at:

Description of the amplitude sweep to determine the boundaries of the linear regime for the linear oscillatory shear measurements; full FMG analysis of stress relaxation results.

## 6. AUTHOR INFORMATION

Corresponding Authors:

Valeriy V. Ginzburg, Department of Chemical Engineering and Materials Science, Michigan State University, East Lansing, Michigan 48224, USA. Email ginzbur7@msu.edu

Eyal Zussman, NanoEngineering Group, Faculty of Mechanical Engineering, Technion-Israel Institute of Technology, Haifa 3200003, Israel. Email meeyal@me.technion.ac.il

Authors:

Agniva Dutta, NanoEngineering Group, Faculty of Mechanical Engineering, Technion-Israel Institute of Technology, Haifa 3200003, Israel

Gleb Vasilyev, NanoEngineering Group, Faculty of Mechanical Engineering, Technion-Israel Institute of Technology, Haifa 3200003, Israel

Notes:

The authors declare no competing financial interest.

## 7. ACKNOWLEDGMENTS

V.G. thanks the Faculty of Mechanical Engineering, Technion, Israel, for hospitality and support during his sabbatical. E.Z. acknowledges the financial support of the Winograd Chair of Fluid Mechanics and Heat Transfer at Technion. We thank Randy Ewoldt (University of Illinois), Gareth McKinley (MIT), Nicholas King (MIT), and Isaac Miranda Valdez (Aalto University) for helpful suggestions.

## 8. REFERENCES

(1) Lee, Y.; Song, W. J.; Sun, J. Y. Hydrogel Soft Robotics. *Materials Today Physics* **2020**, *15*, 100258. https://doi.org/10.1016/j.mtphys.2020.100258.

(2) Naranjo, A.; Martín, C.; López-Díaz, A.; Martín-Pacheco, A.; Rodríguez, A. M.; Patiño, F. J.; Herrero, M. A.; Vázquez, A. S.; Vázquez, E. Autonomous Self-Healing Hydrogel with Anti-Drying Properties and Applications in Soft Robotics. *Appl Mater Today* **2020**, *21*. https://doi.org/10.1016/j.apmt.2020.100806.

(3) Hu, L.; Chee, P. L.; Sugiarto, S.; Yu, Y.; Shi, C.; Yan, R.; Yao, Z.; Shi, X.; Zhi, J.; Kai, D.; Yu, H.; Huang, W. Hydrogel-Based Flexible Electronics. *Advanced Materials* **2023**, *35* (14). https://doi.org/10.1002/adma.202205326.

(4) Ji, C.; Wang, Y.; Qi, Q.; Li, Y.; Cao, L. Electrically Conductive Hydrogels for Flexible Wearable Devices: Materials, Design, and Applications. *Adv Mater Technol* **2025**. https://doi.org/10.1002/admt.202501044.

(5) Dutta, A.; Panda, P.; Das, A.; Ganguly, D.; Chattopadhyay, S.; Banerji, P.; Pradhan, D.; Das, R. K. Intrinsically Freezing-Tolerant, Conductive, and Adhesive Proton Donor-Acceptor Hydrogel for Multifunctional Applications. *ACS Appl Polym Mater* **2022**, *4* (10), 7710–7722. https://doi.org/10.1021/acsapm.2c01285.

(6) Das, S.; Martin, P.; Vasilyev, G.; Nandi, R.; Amdursky, N.; Zussman, E. Processable, Ion-Conducting Hydrogel for Flexible Electronic Devices with Self-Healing Capability. *Macromolecules* **2020**, *53* (24), 11130–11141. https://doi.org/10.1021/acs.macromol.0c02060.

(7) Dutta, A.; Ghosal, K.; Sarkar, K.; Pradhan, D.; Das, R. K. From Ultrastiff to Soft Materials: Exploiting Dynamic Metal–Ligand Cross-Links to Access Polymer Hydrogels Combining Customized Mechanical Performance and Tailorable Functions by Controlling Hydrogel Mechanics. *Chemical Engineering Journal* **2021**, *419*. https://doi.org/10.1016/j.cej.2021.129528.

(8) Chao, D.; Zhou, W.; Xie, F.; Ye, C.; Li, H.; Jaroniec, M.; Qiao, S.-Z. Roadmap for Advanced Aqueous Batteries: From Design of Materials to Applications. *Sci Adv* **2020**, *6* (21), 2375–2548. https://doi.org/10.1126/sciadv.aba4098.

(9) Li, X.; Gong, J. P. Design Principles for Strong and Tough Hydrogels. *Nat Rev Mater* **2024**, *9* (6), 380–398. https://doi.org/10.1038/s41578-024-00672-3.

(10) Dutta, A.; Pandit, S.; Panda, P.; Das, R. K. Competitive (Spatiotemporal) Techniques to Fabricate (Ultra)Stiff Polymer Hydrogels and Their Potential Applications. *ACS Applied Polymer Materials*. American Chemical Society March 28, 2025, pp 3466–3497. https://doi.org/10.1021/acsapm.4c04199.

(11) Okumura, Y.; Ito, K. The Polyrotaxane Gel: A Topological Gel by Figure-of-Eight Cross-Links. *Advanced Materials* **2001**, *13* (7), 485–487. https://doi.org/10.1002/1521-4095(200104)13:7<485::AID-ADMA485>3.0.CO;2-T.

(12) Kim, J.; Zhang, G.; Shi, M.; Suo, Z. Fracture, Fatigue, and Friction of Polymers in Which Entanglements Greatly Outnumber Cross-Links. *Science (1979)* **2021**, *374* (6564), 212–216. https://doi.org/10.1126/science.abg6320.

(13) Sun, W.; Xue, B.; Fan, Q.; Tao, R.; Wang, C.; Wang, X.; Li, Y.; Qin, M.; Wang, W.; Chen, B.; Cao, Y. Molecular Engineering of Metal Coordination Interactions for Strong, Tough, and Fast-Recovery Hydrogels. *Sci Adv* **2020**, *6* (16), 1–12. https://doi.org/10.1126/sciadv.aaz9531.

(14) Suriano, R.; Griffini, G.; Chiari, M.; Levi, M.; Turri, S. Rheological and Mechanical Behavior of Polyacrylamide Hydrogels Chemically Crosslinked with Allyl Agarose for Two-Dimensional Gel Electrophoresis. *J Mech Behav Biomed Mater* **2014**, *30*, 339–346. https://doi.org/10.1016/j.jmbbm.2013.12.006.

(15) Dutta, A.; Vasilyev, G.; Zussman, E. Thermoregulation of Hydrogel Spatiotemporal Mechanics by Exploiting the Folding–Unfolding Characteristics of Globular Proteins. *Biomacromolecules* **2023**, *24* (6), 2575–2586. https://doi.org/10.1021/acs.biomac.3c00073.

(16) Macías-Díaz, J. E. Fractional Calculus—Theory and Applications. *Axioms* **2022**, *11* (2), 43.

(17) Barbero, G.; Evangelista, L. R.; Zola, R. S.; Lenzi, E. K.; Scarfone, A. M. A Brief Review of Fractional Calculus as a Tool for Applications in Physics: Adsorption Phenomena and Electrical Impedance in Complex Fluids. *Fractal and Fractional* **2024**, *8* (7), 369.

(18) Yavuz, M.; Özdemir, N. *Fractional Calculus: New Applications in Understanding Nonlinear Phenomena*; Bentham Science Publishers: Singapore, 2022; Vol. 3.

(19) Dingyü, X.; Lu, B. *Fractional Calculus High-Precision Algorithms and Numerical Implementations*; Springer: Singapore, 2024.

(20) Zayernouri, M.; Wang, L.-L.; Shen, J.; Karniadakis, G. E. *Spectral and Spectral Element Methods for Fractional Ordinary and Partial Differential Equations*; Cambridge University Press: Cambridge, 2024. https://doi.org/DOI: 10.1017/9781108867160.

(21) Jaishankar, A.; McKinley, G. H. Power-Law Rheology in the Bulk and at the Interface: Quasi-Properties and Fractional Constitutive Equations. *Proceedings of the Royal Society A: Mathematical, Physical and Engineering Sciences* **2013**, *469* (2149), 20120284.

(22) Jaishankar, A.; McKinley, G. H. A Fractional K-BKZ Constitutive Formulation for Describing the Nonlinear Rheology of Multiscale Complex Fluids. *J Rheol (N Y N Y)* **2014**, *58* (6), 1751–1788.

(23) Faber, T. J.; Jaishankar, A.; McKinley, G. H. Describing the Firmness, Springiness and Rubberiness of Food Gels Using Fractional Calculus. Part I: Theoretical Framework. *Food Hydrocoll* **2017**, *62*, 325–339.

(24) Keshavarz, B.; Divoux, T.; Manneville, S.; McKinley, G. H. Nonlinear Viscoelasticity and Generalized Failure Criterion for Polymer Gels. *ACS Macro Lett* **2017**, *6* (7), 663–667.

(25) Rathinaraj, J. D. J.; McKinley, G. H.; Keshavarz, B. Incorporating Rheological Nonlinearity into Fractional Calculus Descriptions of Fractal Matter and Multi-Scale Complex Fluids. *Fractal and Fractional* **2021**, *5* (4), 174.

(26) Song, J.; Holten-Andersen, N.; McKinley, G. H. Non-Maxwellian Viscoelastic Stress Relaxations in Soft Matter. *Soft Matter* **2023**, *19* (41), 7885–7906.

(27) Das, M.; Rathinaraj, J. D. J.; Palade, L. I.; McKinley FRS, G. H. Laun's Rule for Predicting the First Normal Stress Coefficient in Complex Fluids: A Comprehensive Investigation Using Fractional Calculus. *Physics of Fluids* **2024**, *36* (1), 13111. https://doi.org/10.1063/5.0179709.

(28) Wagner, C. E.; Barbati, A. C.; Engmann, J.; Burbidge, A. S.; McKinley, G. H. Quantifying the Consistency and Rheology of Liquid Foods Using Fractional Calculus. *Food Hydrocoll* **2017**, *69*, 242–254.

(29) D'Elia, M.; Zayernouri, M.; Gulian, M.; Suzuki, J. *Fractional Modeling in Action: A Survey of Nonlocal Models for Subsurface Transport, Turbulent Flows, and Anomalous Materials*; 2021. https://doi.org/10.13140/RG.2.2.13264.43526.

(30) Tzelepis, D. A.; Khoshnevis, A.; Zayernouri, M.; Ginzburg, V. V. Polyurea–Graphene Nanocomposites—The Influence of Hard-Segment Content and Nanoparticle Loading on Mechanical Properties. *Polymers (Basel)* **2023**, *15* (22), 4434.

(31) Suzuki, J. L.; Zayernouri, M.; Bittencourt, M. L.; Karniadakis, G. E. Fractional-Order Uniaxial Visco-Elasto-Plastic Models for Structural Analysis. *Comput Methods Appl Mech Eng* **2016**, *308*, 443–467.

(32) Tzelepis, D. A.; Suzuki, J.; Su, Y. F.; Wang, Y.; Lim, Y. C.; Zayernouri, M.; Ginzburg, V. V. Experimental and Modeling Studies of IPDI-Based Polyurea Elastomers – The Role of Hard Segment Fraction. *J Appl Polym Sci* **2023**, *140* (10), e53592. https://doi.org/https://doi.org/10.1002/app.53592.

(33) Cambeses-Franco, P.; Rial, R.; Ruso, J. M. Integrating Classical and Fractional Calculus Rheological Models in Developing Hydroxyapatite-Enhanced Hydrogels. *Physics of Fluids* **2024**, *36* (7). https://doi.org/10.1063/5.0213561.

(34) Puente-Córdova, J. G.; Rentería-Baltiérrez, F. Y.; Miranda-Valdez, I. Y. Fractional Rheology as a Tool for Modeling Viscoelasticity in Cellulose-Based Hydrogels. *Cellulose* **2025**, *32* (6), 3619–3632. https://doi.org/10.1007/s10570-025-06468-0.

(35) Lenoch, A.; Schönhoff, M.; Cramer, C. Modelling Viscoelastic Relaxation Mechanisms in Thermorheologically Complex Fe(Iii)-Poly(Acrylic Acid) Hydrogels. *Soft Matter* **2022**, *18* (44), 8467–8475. https://doi.org/10.1039/d2sm01122k.

(36) Das, M.; Waeterloos, J. L.; Clasen, C.; McKinley, G. H. Single Cells Are Compactly and Accurately Described as Fractional Kelvin-Voigt Materials. *Rheol Acta* **2025**. https://doi.org/10.1007/s00397-025-01515-w.

(37) Freund, J. B.; Ewoldt, R. H. Quantitative Rheological Model Selection: Good Fits versus Credible Models Using Bayesian Inference. *J Rheol (N Y N Y)* **2015**, *59* (3), 667–701. https://doi.org/10.1122/1.4915299.

(38) Singh, P. K.; Soulages, J. M.; Ewoldt, R. H. On Fitting Data for Parameter Estimates: Residual Weighting and Data Representation. *Rheol. Acta* **2019**, *58*, 341.

(39) Martinetti, L.; Carey-De La Torre, O.; Schweizer, K. S.; Ewoldt, R. H. Inferring the Nonlinear Mechanisms of a Reversible Network. *Macromolecules* **2018**, *51*, 8772.

(40) Dutta, A.; Maity, P.; Das, R. K.; Zussman, E. Molecular Engineering of Backbone Rotation in an Energy-Dissipative Hydrogel for Combining Ultra-High Stiffness and Toughness. *Mater Horiz* **2025**, *12* (13), 4901–4914. https://doi.org/10.1039/d4mh01717j.

(41) Lin, P.; Ma, S.; Wang, X.; Zhou, F. Molecularly Engineered Dual-Crosslinked Hydrogel with Ultrahigh Mechanical Strength, Toughness, and Good Self-Recovery. *Advanced Materials* **2015**, *27* (12), 2054–2059. https://doi.org/10.1002/adma.201405022.

(42) Rogosin, S.; Mainardi, F. George William Scott Blair--the Pioneer of Factional Calculus in Rheology. *arXiv preprint arXiv:1404.3295* **2014**.

(43) Blair, G. W. S. The Role of Psychophysics in Rheology. *J Colloid Sci* **1947**, *2* (1), 21–32.

(44) Blair, G. W. S.; Veinoglou, B. C.; Caffyn, J. E. Limitations of the Newtonian Time Scale in Relation to Non-Equilibrium Rheological States and a Theory of Quasi-Properties. *Proc R Soc Lond A Math Phys Sci* **1947**, *189* (1016), 69–87.

(45) Winter, H. H.; Chambon, F. Analysis of Linear Viscoelasticity of a Crosslinking Polymer at the Gel Point. *J Rheol (N Y N Y)* **1986**, *30* (2), 367–382.

(46) Chambon, F.; Winter, H. H. Linear Viscoelasticity at the Gel Point of a Crosslinking PDMS with Imbalanced Stoichiometry. *J Rheol (N Y N Y)* **1987**, *31* (8), 683–697.

(47) Larson, R. G. *The Structure and Rheology of Complex Fluids*; Oxford university press New York, 1999; Vol. 150.

(48) Rubinstein, M.; Colby, R. H. *Polymer Physics*; Oxford University Press: Oxford, 2003.

(49) Goswami, M.; Dutta, A.; Kulshreshtha, R.; Vasilyev, G.; Zussman, E.; Volokh, K. Mechanics of Physically Cross-Linked Hydrogels: Experiments and Theoretical Modeling. *Macromolecules* **2025**, *58* (9), 4478–4487. https://doi.org/10.1021/acs.macromol.5c00486.

(50) Winter, H. H. Evolution of Rheology during Chemical Gelation. In *Permanent and Transient Networks*; Springer, 2007; pp 104–110.

(51) Hess, W.; Vilgis, T. A.; Winter, H. H. Dynamical Critical Behavior during Chemical Gelation and Vulcanization. *Macromolecules* **1988**, *21* (8), 2536–2542.

(52) Aime, S.; Cipelletti, L.; Ramos, L. Power Law Viscoelasticity of a Fractal Colloidal Gel. *J Rheol (N Y N Y)* **2018**, *62* (6), 1429–1441.

Supporting Information

# Dual Cross-Linked Hydrogels: Linear Rheology and Fractional Calculus Modeling

Agniva Dutta[a], Valeriy V. Ginzburg[b,*], Gleb Vasilyev[a], and Eyal Zussman[a,*]

[a] NanoEngineering Group, Faculty of Mechanical Engineering, Technion-Israel Institute of Technology, Haifa 3200003, Israel

[b] Department of Chemical Engineering and Materials Science, Michigan State University, East Lansing, Michigan 48224, USA

* Corresponding author emails: ginzbur7@msu.edu (VVG), meeyal@me.technion.ac.il (EZ)

## 1. Amplitude sweeps and the boundaries of the linear region

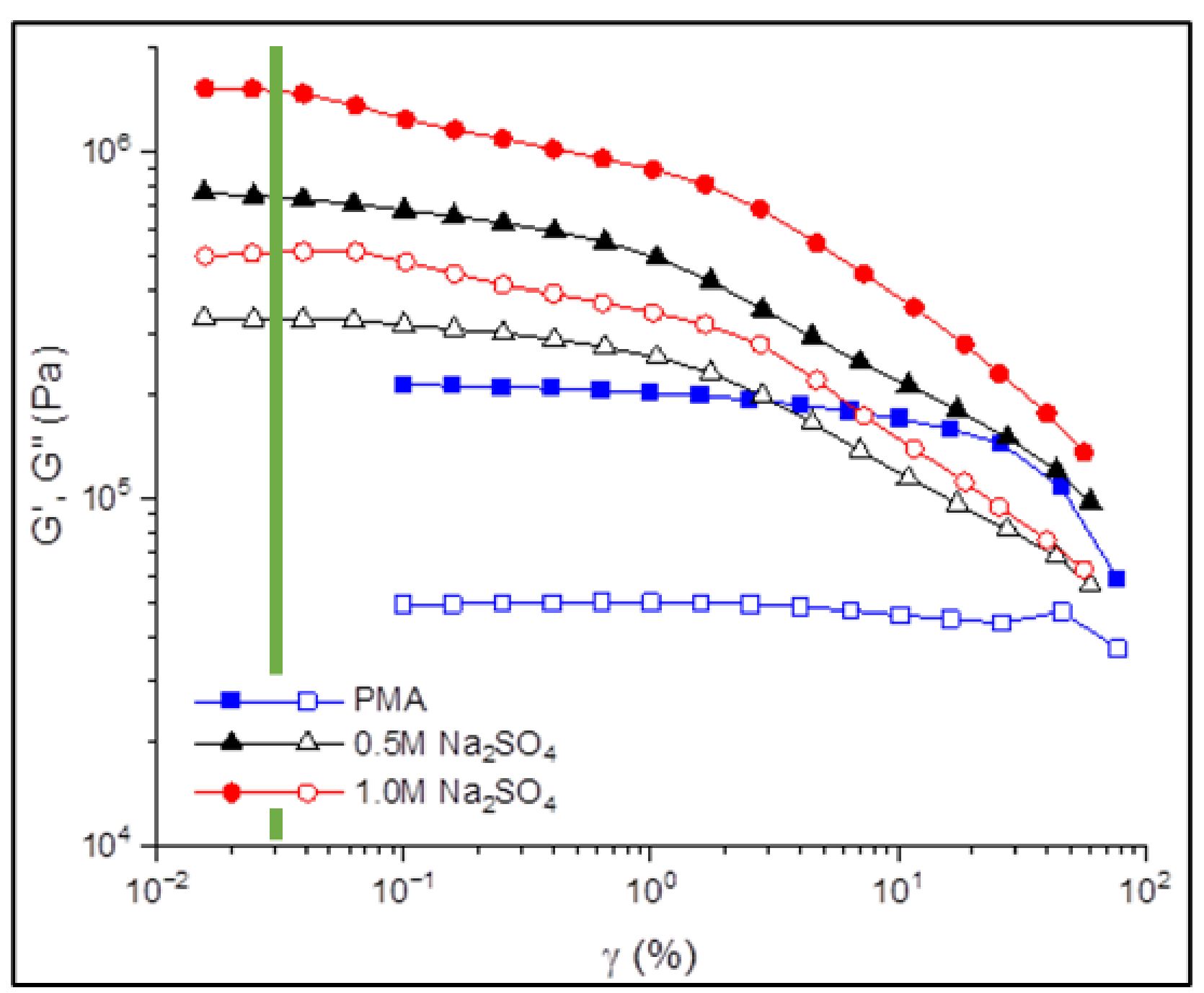

**Figure S1.** The results of the amplitude sweep tests for the hydrogels show a narrowing of the region of linear viscoelastic response of the materials with the increase of the salt concentration. The green line corresponds to the strain amplitude used in the oscillatory shear tests (0.04%).

Figure S1 shows the amplitude sweep (storage and loss moduli) for three gels: PMA (no salt), 0.5M $Na_2SO_4$ PMA, and 1M $Na_2SO_4$ PMA. Based on these curves, we can estimate the boundaries of the linear region (where the G' begins to deviate significantly from the initial plateau) as: PMA (no salt) – 5-10%; 0.5M $Na_2SO_4$ – 0.1-0.3%; and 1M $Na_2SO_4$ – 0.05-0.1%.

## 2. Full FMG analysis of the stress relaxation results

In Figures S2a—e, we plot the full FMG stress relaxation model results, extending the calculation down to 1 microsecond. As a reminder, the experimental setup has a time resolution on the order of 0.1 s. Each of the five figures (corresponding to five gels) depicts the experimental stress relaxation data (squares), the experimental shear modulus determined from the stress-strain test (open squares with dots between them, see Table S1 and Ref. [1]), the FMG stress relaxation curve (solid line), and the zero-time limit of the FMG stress relaxation curve (dashed line). In some cases, the dashed line and the dot-square line or the solid line and the squares lie on top of each other, indicating that they are very close.

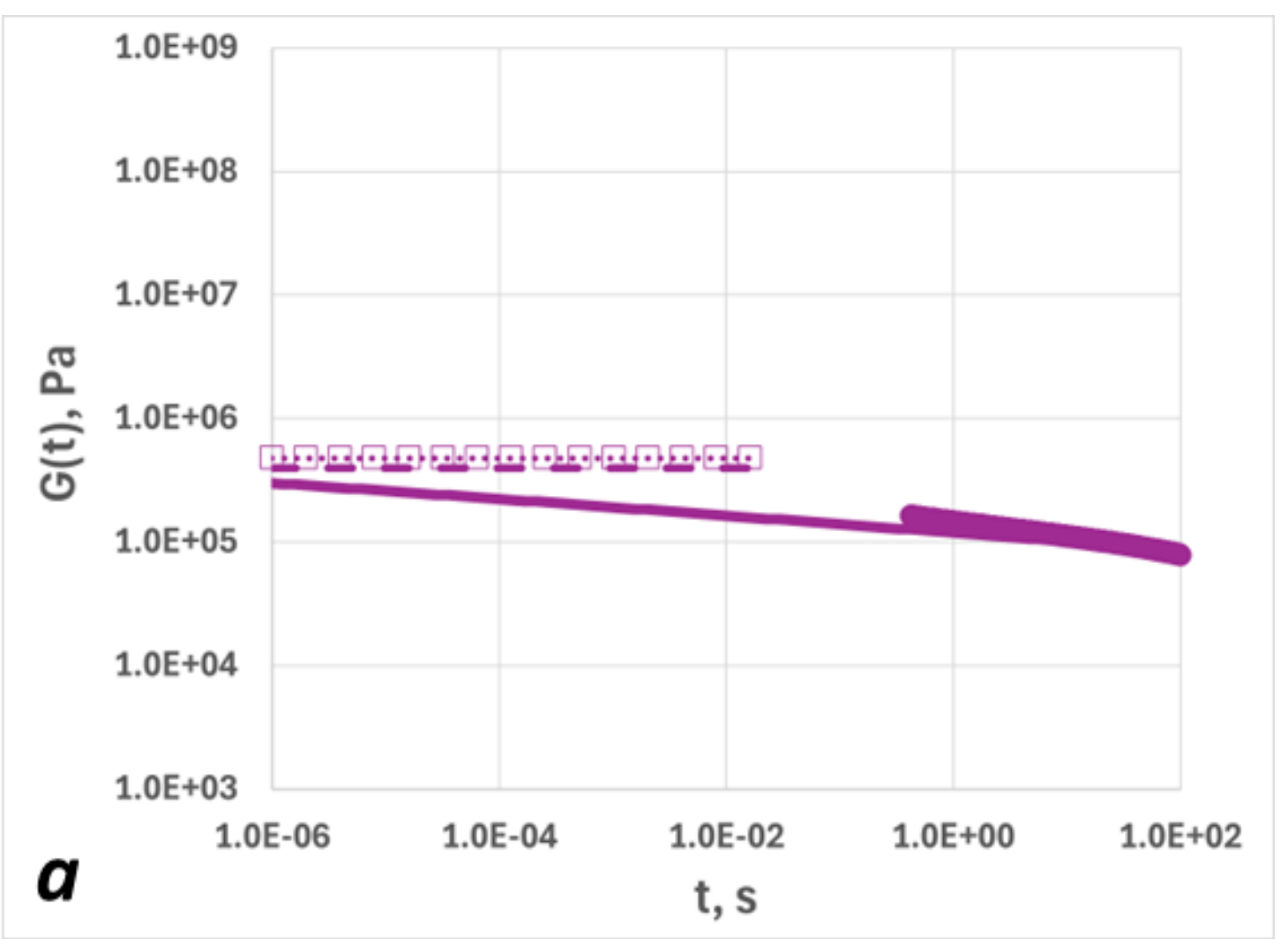

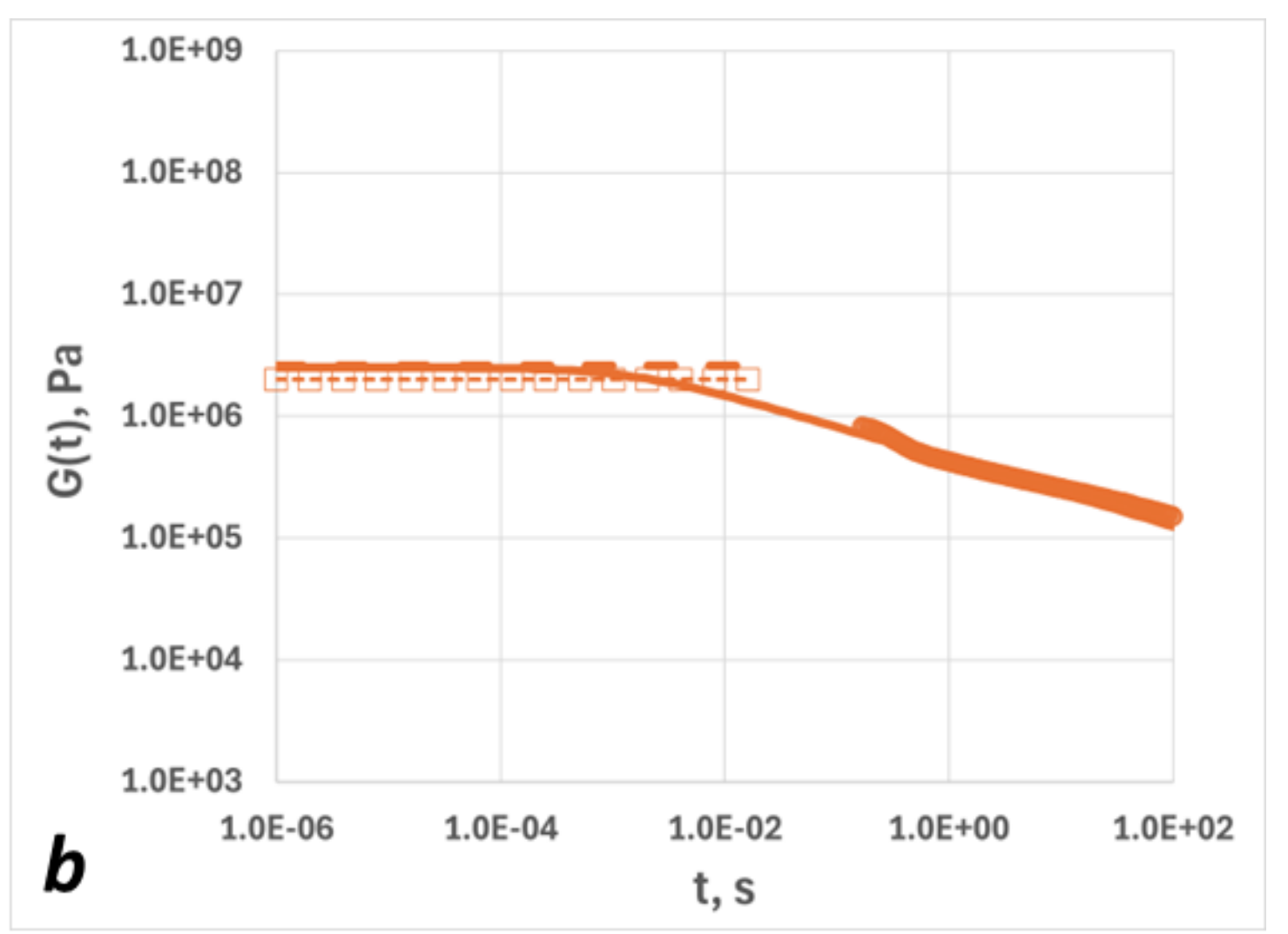

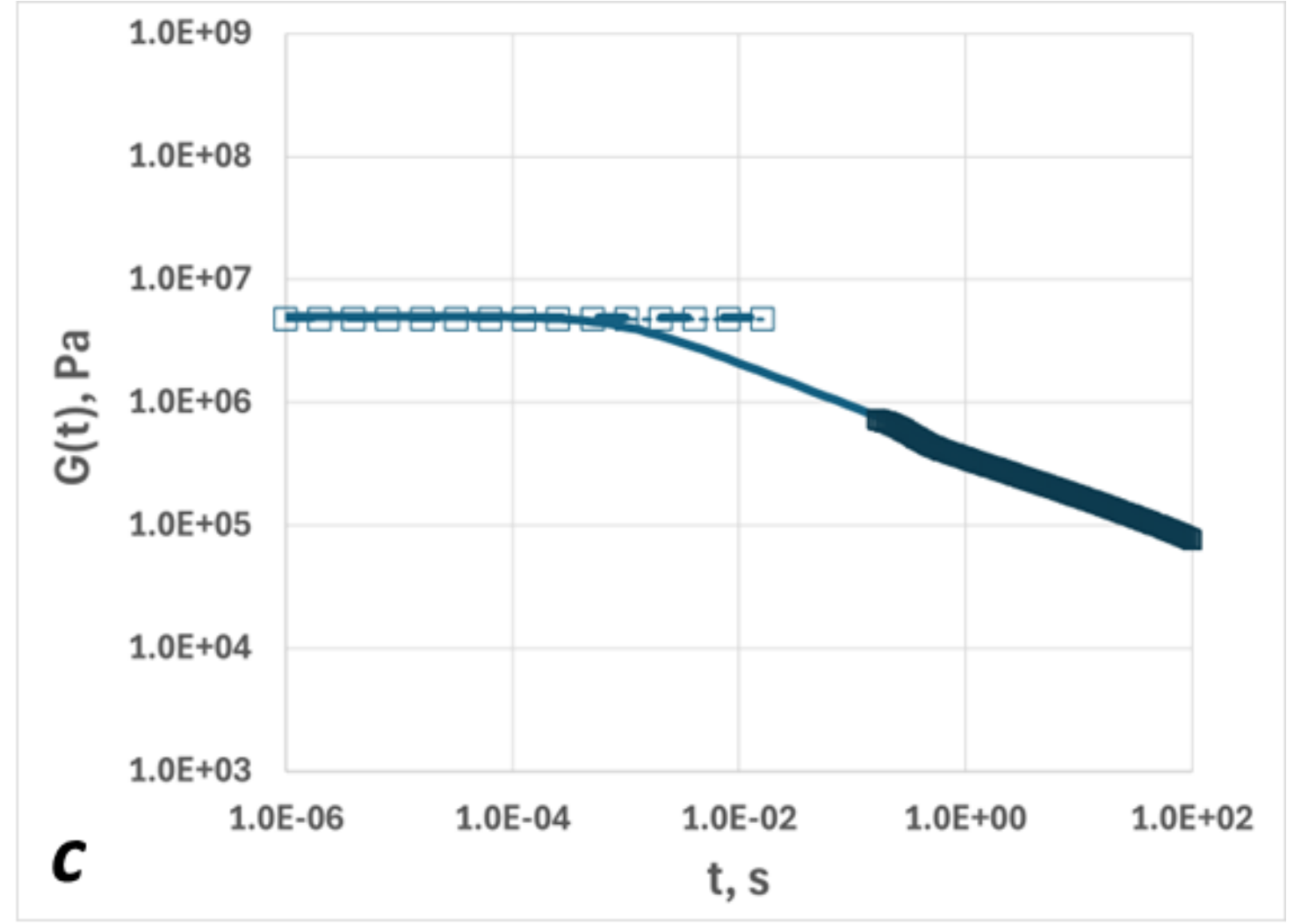

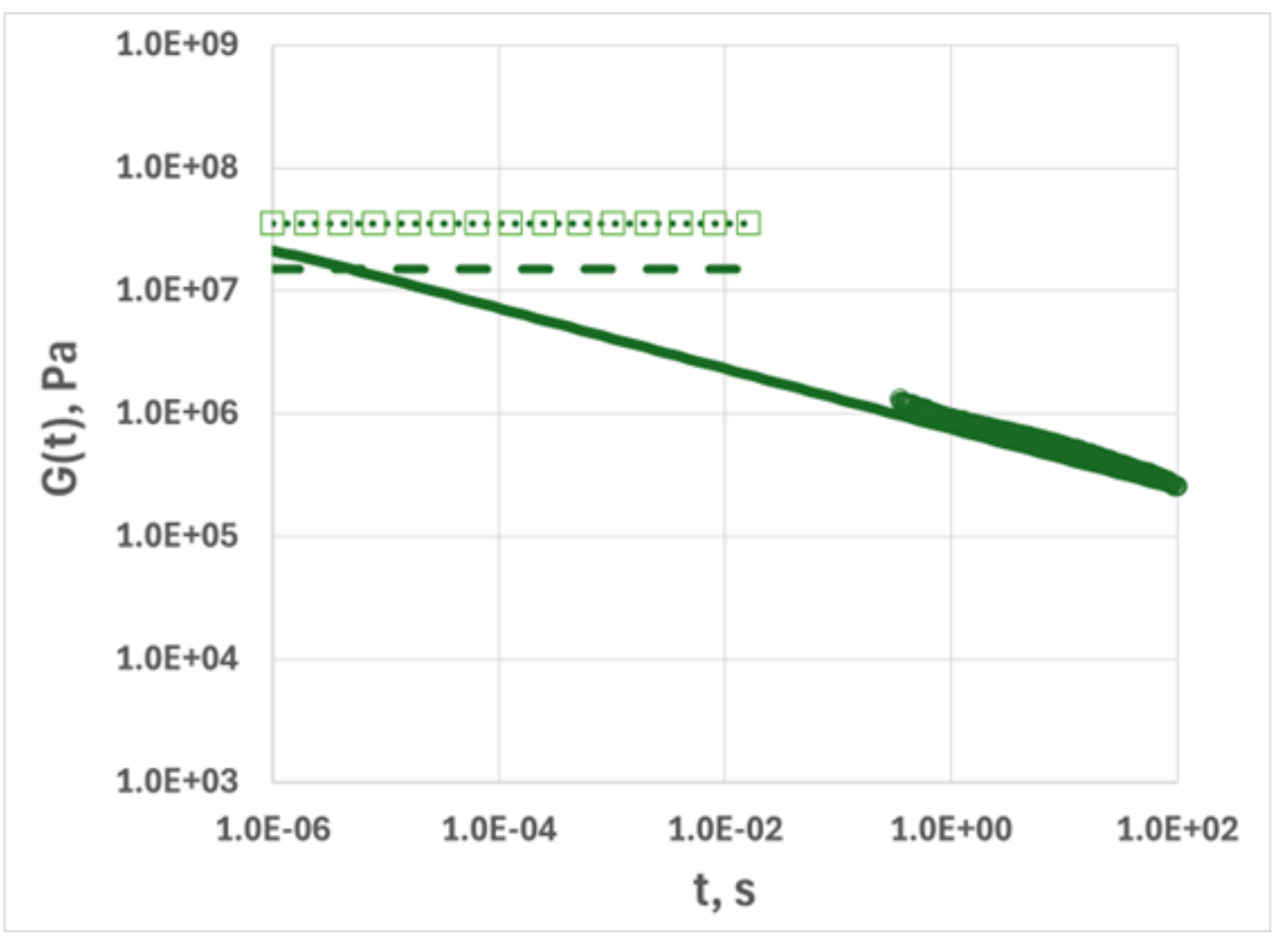

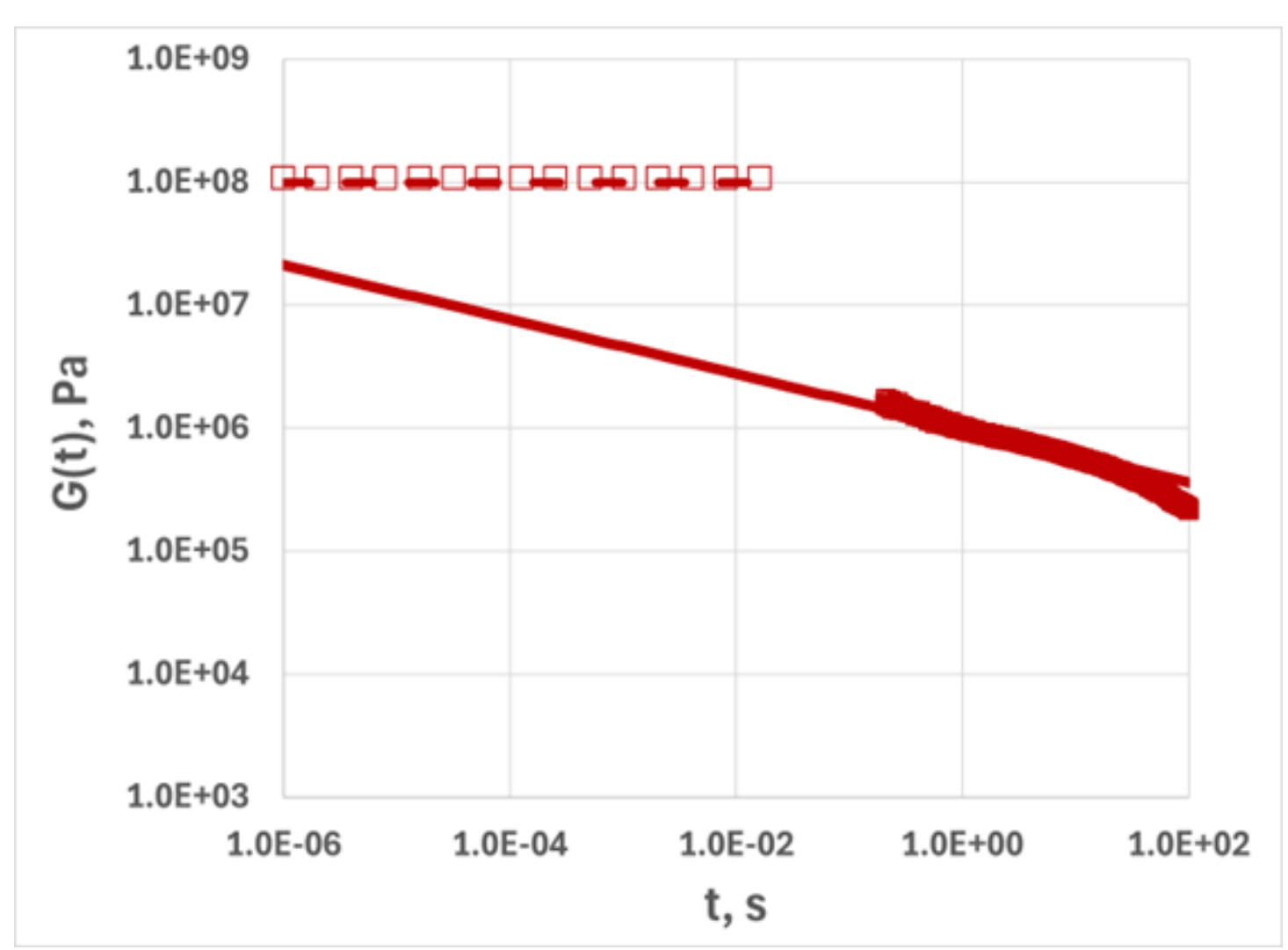

**Figure S2.** The stress relaxation data (squares), FMG fit (solid line), FMG zero-time limit (dashed line), and estimated plateau modulus based on tensile measurements (squares with dots between them). (a) Pure PMA; (b) 0.25M $Na_2SO_4$ PMA; (c) 0.5M $Na_2SO_4$ PMA; (d) 0.75M $Na_2SO_4$ PMA; (e) 1M $Na_2SO_4$ PMA.

**Table S 1**. Experimental stiffness (Young's modulus) of PMA gels with various salt concentrations.

| Concentration (M) | Stiffness (MPa) | Std. Error |
|---|---|---|
| 0 | 1.2 | 0.15 |
| 0.25 | 2.1 | 0.50 |
| 0.5 | 14.5 | 0.64 |
| 0.75 | 107.5 | 11.25 |
| 1 | 325.8 | 37.67 |

[1] A. Dutta, P. Maity, R. K. Das, and E. Zussman, Molecular engineering of backbone rotation in an energy-dissipative hydrogel for combining ultra-high stiffness and toughness, Mater Horiz **12**, 4901 (2025).